\documentclass[letterpaper, 10 pt, conference]{ieeeconf}  

\IEEEoverridecommandlockouts                              

\usepackage{amsmath} 
\usepackage{amssymb}  
\usepackage{bm}

\usepackage{url}            
\usepackage{booktabs}
\usepackage{nicefrac}       
\usepackage{microtype}      

\usepackage{wrapfig}

\usepackage[pdftex]{graphicx}
\graphicspath{{./images/}}
\DeclareGraphicsExtensions{.pdf,.jpg,.png}

\usepackage{mathtools}
\usepackage{setspace}

\newcommand{\lmain}{\text{LSTM}_{main}}
\newcommand{\laux}{\text{LSTM}_{aux}}

\newcommand{\mhyper}{m-HyperLSTM}

\title{Real-Time Workload Classification during Driving using HyperNetworks}

\author{Ruohan Wang and Pierluigi V. Amadori and Yiannis Demiris
\thanks{Authors are with the Personal Robotics Lab, Department of Electrical and Electronic Engineering, Imperial College London, UK {\tt\small \{r.wang16, p.amadori, y.demiris\}@imperial.ac.uk}}. %
}

\usepackage[usenames,dvipsnames,svgnames,table]{xcolor}
\newif\ifdraft\drafttrue
\ifdraft

\else

\fi

\begin{document}

\maketitle
\thispagestyle{empty}
\pagestyle{empty}

\begin{abstract}
Classifying human cognitive states from behavioral and physiological signals is a challenging problem with important applications in robotics. The problem is challenging due to the data variability among individual users, and sensor artefacts. In this work, we propose an end-to-end framework for real-time cognitive workload classification with mixture Hyper Long Short Term Memory Networks (\mhyper{}), a novel variant of HyperNetworks. Evaluating the proposed approach on an eye-gaze pattern dataset collected from simulated driving scenarios of different cognitive demands, we show that the proposed framework outperforms previous baseline methods and achieves 83.9\% precision and 87.8\% recall during test. We also demonstrate the merit of our proposed architecture by showing improved performance over other LSTM-based methods.
\end{abstract}

\section{Introduction}
\label{sec:intro}
Classifying human cognitive states is an important problem with many applications in robotics. In human-robot interaction, the ability to predict human intentions enables robots to perform socially compliant navigation and collaborate with humans~\cite{Demiris2007, kretzschmar2013, koppula2016, mainprice2013human}. For intelligent vehicles, intention or distraction prediction allows the systems to alert users before potentially dangerous maneuvers~\cite{liang2007,jain2016,wollmer2011}. Further, casting driving assistance as a problem of human-in-the-loop control, users' cognitive states provide input for deriving control policies to manage user interfaces and take over control if necessary~\cite{healey2005, Lam2015, driggs2015, carlson2012}.


\begin{figure}[t]
\centering
\includegraphics[width=0.40\textwidth]{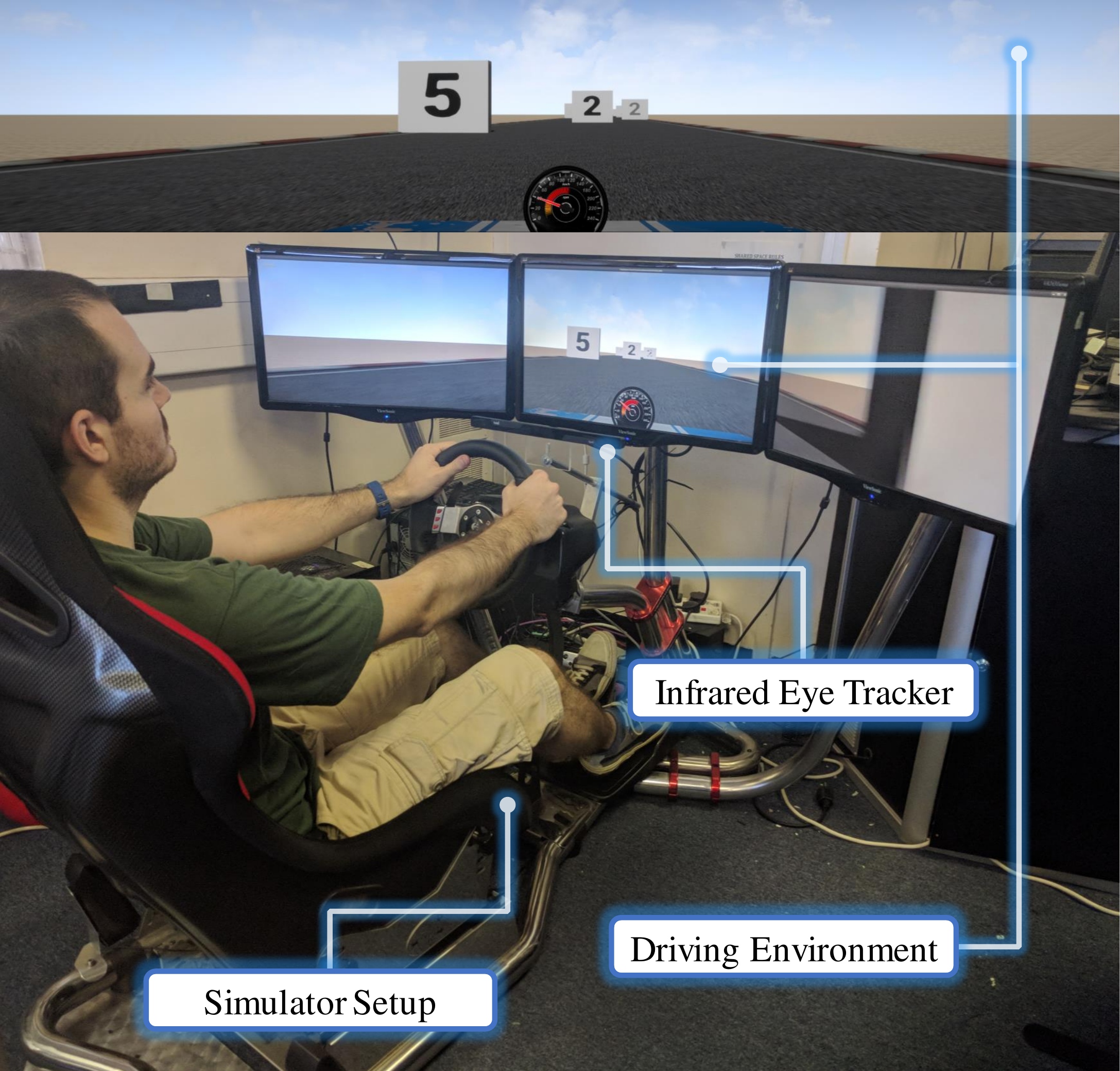}
\caption{Top: Simulated driving environment. Numbered obstacles are placed in three lanes. Participants drive along the road to avoid obstacles, and perform mental tasks, while their performance data, and their instantaneous gaze locations are recorded. Bottom: Simulator physical setup.}
\label{fig:setup}
\end{figure}

Previous studies show that physiological and behavioral signals correlate with cognitive states. For instance, \cite{jain2016} used head movements to predict intention in driving, while \cite{healey2005} showed real-time quantitative correlation between stress and physiological signals including Electrocardiogram (ECG), skin conductance, and respiration in different individuals. Common challenges demonstrated in those studies are data variability among individuals and sensor artefacts. Given the temporal nature of physiological data, heavy features engineering are commonly employed to improve data quality and summarize the data into fixed-size features suitable for classification algorithms such as logistic regression and Support Vector Machine (SVM). However, it is desirable for a model to 1) automatically learn feature representations from data to reduce manual feature engineering, and 2) be sufficiently flexible to tackle data variability.


Towards the goals stated above, we propose a framework for real-time cognitive workload classification using mixture Hyper Long Short Term Memory Networks (\mhyper{}), a novel variant of HyperNetworks~\cite{Ha2017} based on LSTM~\cite{hoch1997}. HyperLSTM is a class of HyperNetworks wherein the model parameters adapt based on the input. Our choice of the model is motivated by the hypothesis that the adaptive nature of HyperNetworks can be exploited to tackle data variability while LSTMs are known for their ability to capture long range dependency and learning useful feature representations from data~\cite{hoch1997}. We formulate the classification task as learning a sequence-to-sequence mapping. 

\begin{figure}[t]
\centering
\includegraphics[width=.48\textwidth]{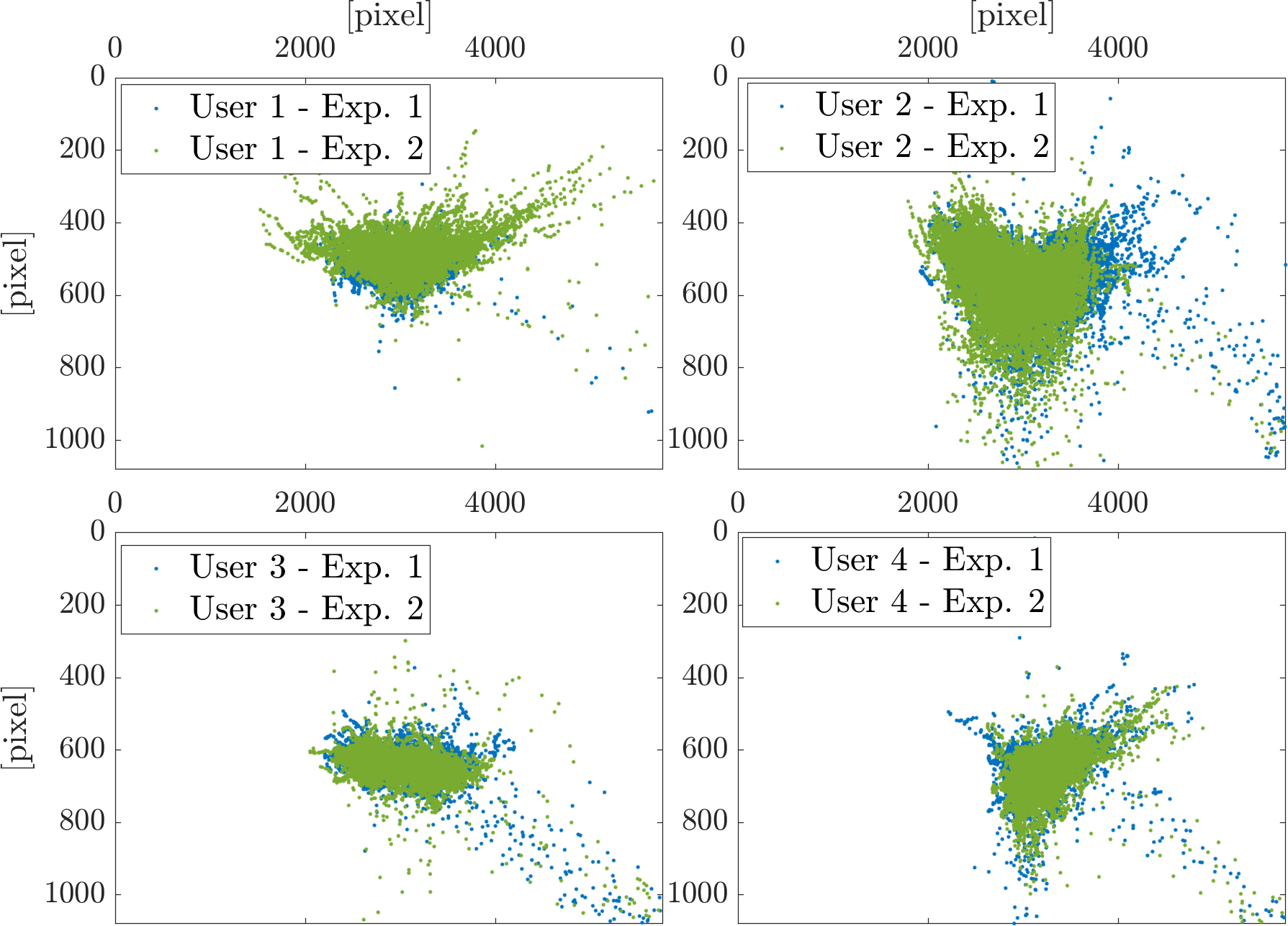}
\caption{Scatter plot of instantaneous gaze locations of four participants during low (blue) and high (green) workload situations. The plots highlight that the dataset is challenging as eye gaze patterns differ among individuals. Best viewed in color.}
\label{fig:scatter}
\end{figure}

We collect an eye-gaze location dataset from simulated driving whereby 20 participants complete tasks of different cognitive demands. We then evaluate our proposed approach on the dataset for binary classification of cognitive workload levels (low and high). The classification is challenging as the dataset is both noisy and exhibits varying visual scanning patterns across individuals, as shown in Fig. \ref{fig:scatter}. Similar to \cite{liang2007}, we choose eye gaze as input because eye tracking is less intrusive compared to skin conductance or ECG tracking, and readily available through consumer products (e.g. cameras in natural environments ~\cite{Fischeretal2018}  or smart phones~\cite{krafka2016}, and has wider applicability beyond driving. We stress that the proposed framework is generic to different sensor modalities and multi-class classification, and we intend to explore sensor fusion and more fine-grained cognitive states classification in future works, including the usage of skin conductance, ECG data and extending to multi-class classifications with different workload levels.

We report experiment results comparing \mhyper{} with different baseline models, including state-of-the-art sequence models such as HyperLSTM~\cite{Ha2017}, and LSTM~\cite{jain2016, wollmer2011}, as well as classical models such as SVM~\cite{solovey2014, liang2007}, and logistic regression~\cite{Georgiou2017,solovey2014}. The proposed approach outperforms the baselines and achieves a 83.9\% precision and 87.8\% recall on the test sets. Improved performance over HyperLSTM and LSTM validates the efficacy of the proposed architecture in tackling the variability in the dataset. Key contributions of the paper are:
\begin{itemize}
\item We introduce \mhyper{} for real-time cognitive workload estimation. The architecture jointly learns feature representations and adapts itself based on input. Our contribution is a novel weight generation scheme inspired by the idea of mixture models, aimed at tackling data variability and improving generalization performance.
\item We evaluate the performance of the proposed model against baseline approaches using multiple evaluation metrics, including F1-score, precision, and recall.
\item We validate the proposed architecture ability to handle data variability in simulated driving tasks, in comparison with LSTM variants commonly used for sequence modeling.
\end{itemize}

\section{Related Work}
Our work is related to previous works on cognitive states classification, and on Recurrent Neural Networks (RNNs) for sequence prediction.

Using physiological signals for cognitive workload classification presents multiple challenges. Sophisticated engineering is often required to improve data quality and extract useful features from raw senor signals (e.g. \cite{healey2005,liang2007,solovey2014}. On the other hand, the extent of statistical correlations between cognitive workload and physiological signals can vary significantly among experiment participants \cite{healey2005}. One possible solution to data variability is personalized models, as seen in \cite{ferre2014, Georgiou2017}. However, this approach may become impractical as data collection and model training are required for every new user. Similar to \cite{jain2016,wollmer2011}, our proposed model aims to learn feature representations directly from data. In addition, the adaptive nature of the proposed model directly tackle data variability. Our experiments demonstrate that adaptive models outperform the static ones, hence suggesting a viable generic technique for tackling data variability found in physiological signals.

LSTM Networks~\cite{hoch1997} and HyperNetworks~\cite{Ha2017} are the core components of the proposed model. LSTM Networks were designed to capture long-range dependency within input sequences, and have been shown effective across various sequence modeling tasks, including natural language processing~\cite{sutskever2014sequence} and robotic perception~\cite{finn2016}. HyperNetworks refer to the general approach of generating the weights of a network by another network. Specifically, HyperLSTM extends LSTM by using an auxiliary LSTM to dynamically generate the weights for the main LSTM at each time step, thus enabling the main LSTM to adapt itself based on the input. HyperLSTM has been shown to outperform LSTM in language modeling, handwriting generation and neural machine translation~\cite{Ha2017}. We introduce \mhyper{} inspired by the idea of mixture models, to further exploit the dynamic property of HyperNetworks for cognitive workload classification.

\section{Method}
We cast the task of cognitive workload classification as supervised learning. Given a dataset $\{(\bm{x}_1, \bm{x}_2, ..., \bm{x}_T)_j,\bm{y}_j\}_{j=1}^N$ where $\bm{x}_t$ denotes physiological signals at time $t$, $y$ the target workload level for the sequence, and T the sequence length, we aim to learn a model $\theta$ that maximizes the probability $p(\bm{y} | \bm{x}_{1,..,T})$. Instead of relying a single label, we follow \cite{jain2016} to train our model using the following loss function in a sequence-to-sequence manner
\begin{equation}
\label{eq:loss}
loss=\sum_{j=1}^N\sum_{t=1}^T -e^{(t-T)}\text{log}p(y_t^k|\bm{x}_{<t})
\end{equation}
where $\bm{x}_{<t}$ denotes the subsequence $\bm{x}_{1,..,t}$, and $y_t^k$ the probability of ground truth label computed by the model at time $t$. In addition to encouraging the network to fix early mistakes and reducing the possibility of over fitting when the current context
is insufficient for classification~\cite{jain2016}, the loss function also reduces model variance (i.e., changing predicted label between time steps). We implement our model as \mhyper{} described in Section \ref{sec:hyper}.

\subsection{Long Short-Term Memory Networks}
LSTM is a RNN that implements a memory cell to maintain contextual information over time, and thus captures long-range dependencies in data sequences~\cite{hoch1997}. Given an input sequence $\bm{x}_1, \bm{x}_2,..., \bm{x}_T$, LSTM maps the input sequence to a sequence of hidden states $\bm{h}_1, \bm{h}_2,..., \bm{h}_T$ via the following updates:
\begin{align}
		\bm{i}_t &= \sigma(\bm{W}^i\bm{h}_{t-1} + \bm{I}^i\bm{x}_t+\bm{b}_i)\\
        \bm{f}_t &= \sigma(\bm{W}^f\bm{h}_{t-1} + \bm{I}^f\bm{x}_t+\bm{b}_f)\\
        \bm{o}_t &= \sigma(\bm{W}^o\bm{h}_{t-1} + \bm{I}^o\bm{x}_t+\bm{b}_o)\\
		\bm{c}_t &= \bm{f}_t \odot \bm{c}_{t-1} + \bm{i}_t \odot \tanh(\bm{W}^c\bm{h}_{t-1} + \bm{I}^c\bm{x}_t+\bm{b}_c)\label{eq:lstm_mem}\\
		\bm{h}_t &= \bm{o}_t \odot \tanh(\bm{c}_{t}),
\end{align}
where $\odot$ denotes the element-wise product, $\sigma$ and $\tanh$ the element-wise sigmoid function and hyperbolic tangent function respectively. $\bm{W}^*$, $\bm{I}^*$, and $\bm{b}_*$ are parameters to be learned, with $*$ represents one of $\{i, c, f, o\}$ gates. Eq. \ref{eq:lstm_mem} shows that the memory $\bm{c}_t$ of LSTM selectively carries information from the previous time step by controlling what to remember via the forget gate $\bm{f}_t$. The LSTM defined above is similar to the architecture of \cite{graves2013generating} but without peep-hole connections. For simplicity, we use the following shorthand for LSTM updates:
\begin{equation*}
(\bm{h}_t, \bm{c}_t) = \text{LSTM}(\bm{h}_{t-1}, \bm{c}_{t-1}, \bm{x}_t).
\end{equation*}


\subsection{HyperLSTM}
\label{sec:hyper}
\begin{figure}[t]
\centering
\includegraphics[width=.48\textwidth,trim=2.2cm .5cm .5cm .5cm, clip]{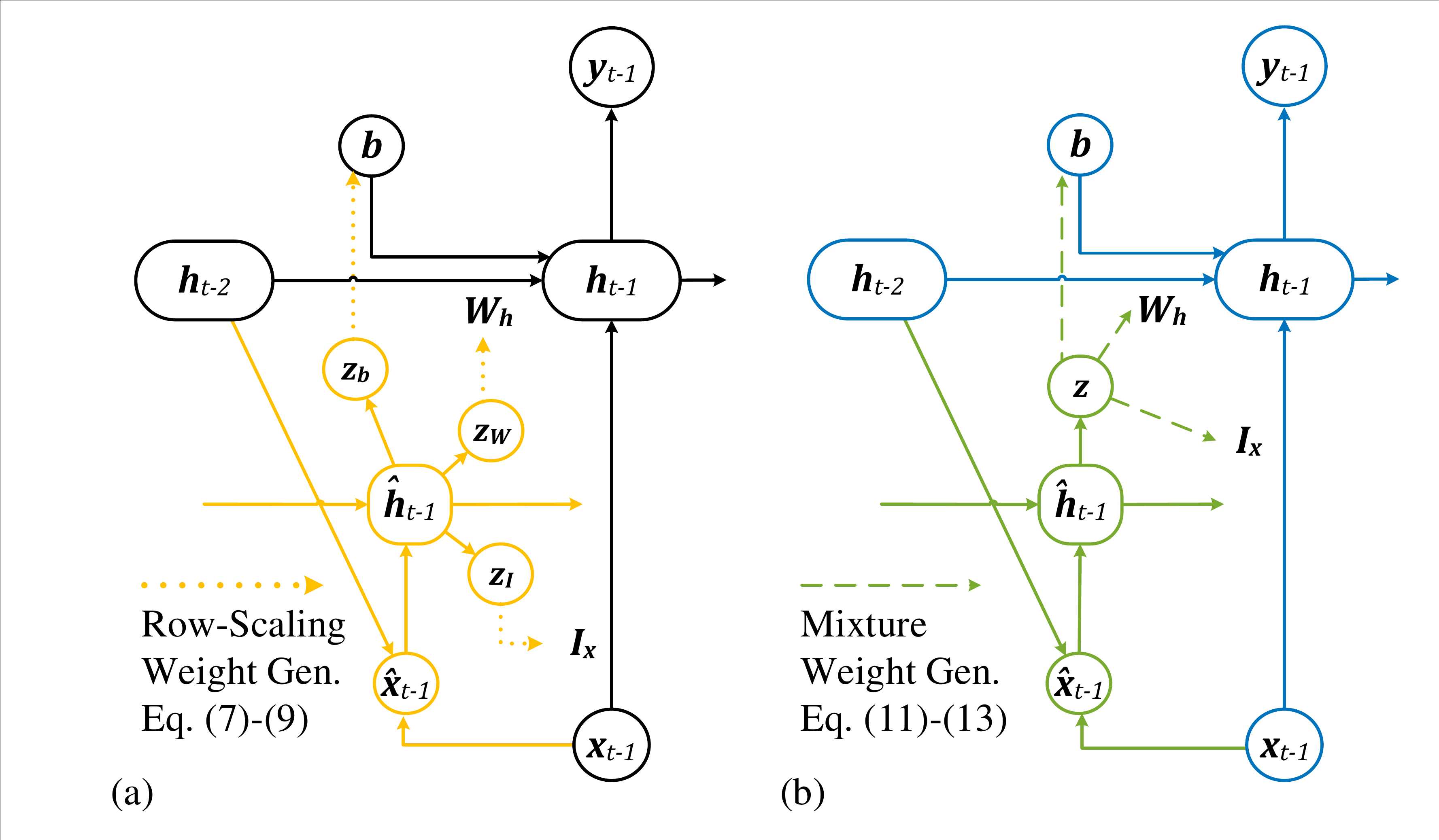}
\caption{Comparison between the original HyperLSTM (left) and the proposed m-HyperLSTM architecture (right). Key differences between the two are the weight generation schemes and the associated parameter sharing. Here, $x_t$ represents input at time $t$, $h_t$ denotes learned features/hidden states, and $y_t$ identifies prediction output.}
\label{fig:hyper}
\end{figure}

HyperNetworks is a family of network architectures that generates the weights of a network via another network, and has achieved state-of-the-art performance in various sequence modeling tasks~\cite{Ha2017}. In particular, HyperLSTM extends LSTM by using an auxiliary LSTM ($\laux$) to dynamically generate the weights of a main LSTM ($\lmain$), shown in Fig. \ref{fig:hyper}a. Following the update for $\laux$ in \cite{Ha2017}, we have
\begin{align*}
\bm{\hat{x}}_t &=
\begin{pmatrix}
  \bm{x}_t\\
  \bm{h}_{t-1}
\end{pmatrix}\\
(\bm{\hat{h}}_t, \bm{\hat{c}}_t) &= \laux(\bm{\hat{h}}_{t-1}, \bm{\hat{c}}_{t-1}, \bm{\hat{x}}_t),
\end{align*}
where the input $\bm{\hat{x}}_t$ to $\laux$ is the concatenation of the current input $\bm{x}_t$ and the hidden state $\bm{h}_{t-1}$ from $\lmain$. HyperLSTM then parametrizes the weights of $\lmain$ at time $t$ as a function of $\bm{\hat{h}}_{t}$. For $\bm{W}^*_t$, it is defined as
\begin{align}
\bm{z}^*_h &= \bm{W}^*_{hz}\bm{\hat{h}}_{t}+\bm{b}^*_h\\
\bm{d}^*_h &= \bm{W}^*_{hd}\bm{z}^*_h\\
\bm{W}^*_{t} &= \begin{pmatrix}
  (\bm{d}^*_h)_1\bm{W}^*_{(1)}\\
  (\bm{d}^*_h)_2\bm{W}^*_{(2)}\\
  ...\\
  (\bm{d}^*_h)_{N_h}\bm{W}^*_{(N_h)}\\
\end{pmatrix}_,
\end{align}
where $\bm{W}^*_{(j)}$ denotes the $j$-th row of $\bm{W}^*$, and $\bm{W}^*_{hz}$, $\bm{b}^*_h$ and $\bm{W}^*_{hd}$ parameters to be learned. Both $\bm{I}^*_{t}$ and $\bm{b}^*_{t}$ follow the same update rule, omitted here for brevity. For further details on HyperLSTM, we refer the readers to \cite{Ha2017}.


\section{\mhyper{}}
Many other mappings from current contexts to network weights are possible. We present a novel scheme for weights generation, inspired by the idea of mixture models, shown in Fig. \ref{fig:hyper}b. The scheme mixes $N_z$ LSTMs before the nonlinearity with their activation strengths (from 0 to 1) determined by the current context. Analytically, we formulate the update rule as
\begin{align}
\bm{z} &= \sigma(\bm{W}^z\bm{\hat{h}}_{t}+\bm{b}^z)\label{eq:hyper_first}\\
\bm{W}^*_{t} & = <\bm{W}^*_{z}, \bm{z}>\\
\bm{I}^*_{t} & = <\bm{I}^*_{z}, \bm{z}>\\
\bm{b}^*_{t} & = <\bm{b}^*_{z}, \bm{z}>\label{eq:hyper_last},
\end{align}
where $\bm{W}^*_z \in \mathbb{R}^{N_h\times N_h\times N_z}$, $\bm{I}^*_z \in \mathbb{R}^{N_h\times N_x\times N_z}$, $\bm{b}^*_z \in \mathbb{R}^{N_h\times N_z}$. $N_h$, $N_x$ and $N_z$ denote the dimensions of $\bm{h}_{t}$, $\bm{x}_t$ and $\bm{z}$ respectively. $<\circ, \circ>$ denotes the dot product.

The key differences between the proposed weight generation scheme and the original scheme are 1) parameter allocation and 2) increased regularization. Given a fixed parameter budget, our model trade-offs the size of hidden states for more expressive weight generation, while the original model does the opposite. In addition, the proposed scheme is more flexible as it may learn up to $N_z$ components by turning on and off each element of $\bm{z}$ independently, which helps to prevent overfitting. If only a single element of $\bm{z}$ is turned on at all time steps, our model reduces to a standard LSTM Network. Further, our scheme shares $\bm{z}$ for all weights generation to improve regularization. We found that \mhyper{} outperforms the original in the experiments, suggesting that expressive weight generation and additional regularization contribute to the improved performance.


\subsection{Network Architecture and Training Procedure}
\label{sec:train}
Given an input sequence $\{\bm{x}_1, \bm{x}_2,..., \bm{x}_T\}$, we use the proposed architecture to map the input sequence to a sequence of hidden states $\{\bm{h}_1, \bm{h}_2,..., \bm{h}_T\}$. We then project the hidden states with a fully-connected layer with Rectified Linear Unit (ReLU) nonlinearity, followed by a softmax layer to predict the probability for each possible label.
\begin{equation*}
y_t=\text{softmax}(\bm{W}_2\text{ReLU}(\bm{W}_1\bm{h}_t+\bm{b}_1)+\bm{b}_2).
\end{equation*}

To stabilize hidden state dynamics, we apply layer normalization as suggested in \cite{Ha2017}. To improve generalization, we employ L2 regularization and a label smoothing technique~\cite{szegedy2016rethinking}. The label smoothing technique penalizes overconfident predictions by assigning $1-\epsilon + \frac{\epsilon}{K}$ probability to the correct label, and $\frac{\epsilon}{K}$ to all other labels, where $\epsilon$ is a tunable hyper parameter, and $K$ the number of possible labels. Label smoothing replaces $log\text{ } p(y^k_t|\bm{x}_{<t})$ in Eq. \ref{eq:loss} with cross-entropy $H(p, y_t)$ where $p$ is the smoothed ground truth distribution. Label smoothing naturally fits with cognitive workload classification as there is inherent uncertainty in the ground truth labels, given that cognitive workload is not directly observable~\cite{gopher1986}. All models are trained with Adam~\cite{adam} with a fixed learning rate of 0.0001. We set $\epsilon=0.2$ as recommended in \cite{szegedy2016rethinking}.

\section{Experiments}
We evaluate the proposed approach on the real-time classification of cognitive workload using eye gaze patterns. We detail in the following sections the experimental procedures for collecting the gaze location dataset of the participants under different cognitive workloads. We evaluate our architecture on the collected dataset and compare it to baseline methods, including LSTM~\cite{jain2016,wollmer2011} HyperLSTM~\cite{Ha2017}, SVM~\cite{solovey2014,liang2007} and logistic regression~\cite{solovey2014,Georgiou2017}. We aim to address the following questions:
\begin{itemize}
\item Is \mhyper{} capable of learning useful feature representations from eye gaze patterns and classifying cognitive workload across individuals in driving scenarios?
\item How does \mhyper{} compare to the state-of-the-art sequence models as well as classical methods in terms of classification performance?
\end{itemize}

\subsection{Participants}
Twenty participants (12 males, 8 females, mean age 24.3, standard deviation 3.2) with normal or corrected to normal vision consented to participate in the experiment. After a brief introduction to the experiment and calibration procedure, participants were given a trial period to familiarize themselves with the simulator environment before the actual experiment.

\subsection{Setup}
A realistic driving simulation was set up for the experiment (Fig. \ref{fig:setup}). The environment comprised of monitors, a physical simulator, and a remote eye tracker, mounted above the steering wheel. We developed a customized simulated driving environment based on the Unreal Engine (Fig. \ref{fig:setup}).

\subsection{Experimental Procedure}
Since cognitive workload is not directly observable~\cite{gopher1986}, we follow previous approaches~\cite{liang2007,healey2005,solovey2014,carlson2012} to modulate the cognitive workload experienced by the participants by varying task difficulties using a validated experimental approach for workload generation. The experiment includes two scenarios with different workload levels (low and high), and therefore binary ground truth labels. Though only a coarse classification of cognitive workload is considered in this work, the information is nevertheless an important input for assistive robotics as demonstrated in \cite{Lam2015,driggs2015,carlson2012}. We also intend to explore more fine-grained classification in future works.

For both low and high workload scenarios, the primary objective is to drive along a straight road and avoid stationary rectangular obstacles. The obstacles are numbered 0 through 9 and placed at a regular interval (Fig. \ref{fig:setup}). Participants are asked to maintain their speed between 120 and 130 km/h to ensure a consistent workload level throughout the scenarios. Participants were asked to repeat a scenario if their driving speed deviated from the specified range by 10km/h. The road is divided into three lanes and obstacles are randomly placed at one of three lanes. Obstacles are designed to block an entire lane, so that participants must steer to avoid them. Furthermore, to ensure that a lane is not free of obstacles for extended periods of time, thus reducing primary task difficulty, we employ a custom-defined discrete distribution for obstacle placement. Consider $c_i$ as the distance between the current obstacle location and the previous obstacle location in lane $i$, we define the obstacle placement probability distribution in lane $i$ as
\begin{equation*}
p(i)=\frac{e^{c_i/\text{IntervalSize}}}
{\sum_ie^{c_i/\text{IntervalSize}}}_.
\end{equation*}
where IntervalSize represents the distance between two adjacent obstacles. This ensures that a lane would almost certainly be blocked if the lane has not been chosen for the previous few obstacles.

We employ a visual "$n$-back" task~\cite{nback} as the secondary objective for controlling the workload level of the participants. The task induces different levels of cognitive workload by varying the amount of information that participants need to memorize in their working memory. This approach has been validated in previous studies to provide a consistent level of cognitive workload~\cite{nback,mehler2012sensitivity,reimer2012field}. In our experiment, low workload scenario is associated with a 0-back task (i.e., no memorization required), wherein participants are simply required to determine the parity of the number on the nearest obstacle ahead, and press a corresponding button located on the steering wheel. In the high workload scenario, a 1-back task is employed, so that participants need to recall the parity of the number on the previous obstacle and, as they drive past a new obstacle, press the corresponding button. It is important to stress that the only difference between the two scenarios is the additional cognitive workload stemming from the memorization of numbers. This is pivotal for mitigating the risk of the model classifying other variables, such as additional visual stimuli rather than the cognitive workload.


\subsection{Data Collection and Pre-processing}
We collected instantaneous gaze locations in the reference plane of the center monitor at 60 Hz. The data is recorded in the format of \{\textit{timestamp, x-coordinate, y-coordinate}\}. For each sample, we augment the data with the following attributes: distance from the previous sample (horizontal distance, vertical distance and overall) and the instantaneous speed from previous sample (horizontal speed, vertical speed and overall). In total, each time step contains 8 attributes \{\textit{x-coordinate, y-coordinate, x-distance, y-distance, distance, x-speed, y-speed, speed}\}.

For logistic regression and SVM, we reduce a temporal sequence of attributes into a fixed-size feature vector by capturing the central tendencies, variability, and extremes of each attribute. These features include mean, standard deviation, median, 25th and 75th percentiles, maximum, minimum and range over a window size of $t_w$ seconds for each attribute, resulting in a total of $8\times 8=64$ features. The window size determines the amount of context available for classification and directly impacts the real-timeliness of the method. For LSTM-based models, the input sequence consists of $t_w$ steps for the same window size, with the input for each time step being the features defined above across a 1-second window. All features are normalized to the interval $[0, 1]$ and we uniformly sample the input sequences using a sliding window with 90\% overlap to generate training samples for all the models.

\subsection{Evaluation Setup}
We follow an evaluation framework similar to \cite{lipton201,jain2016}. The evaluation metrics include precision, recall, and F1-score. We train on 80\% of data, setting aside 10\% each for validation and testing using uniformly random splits. We use the validation set to select the model with lowest validation loss within 50 epochs of training, and the decision threshold that maximizes the F1-score. For each model, we report the mean and the standard deviation for each metrics over five randomly sampled and non-overlapping test sets.

For SVM, we use a simple grid search to determine the best hyper parameters ($C=5, \gamma=0.01$). For logistic regression, L2 regularization is used. For all LSTM-based methods, the training procedures and the usage of regularization techniques are identical, as described in \ref{sec:train}. We choose the network sizes for all LSTM-based models such that each model has approximately the same number of parameters and thus similar model capacity. Specifically, we consider a LSTM with a hidden state size of 100. For the original HyperLSTM, we consider a $\lmain$ with hidden state size of 75, a $\text{LSTM}_{aux}$ with hidden state size of 16, and $N_z=4$. For the proposed model, we use a hidden state size of 32 for $\lmain$ and all other settings are identical to those of the original HyperLSTM.

\section{Results}
We present the classification performance of all evaluated models in Table \ref{tab:res}, for $t_w=5s, 10s, 20s$ respectively. 
\setlength\tabcolsep{4pt}
\begin{table*}[t]\footnotesize
\caption{Classification Performance on Cognitive Workload using Gaze Location Sequence}
\label{tab:res}
\begin{center}
\begin{tabular}{c|ccc|ccc|ccc}
\hline
& \multicolumn{3}{c|}{5s} & \multicolumn{3}{c|}{10s} & \multicolumn{3}{c}{20s}\\
\hline
& F1 & Pr (\%) & Re (\%) & F1 & Pr (\%) & Re (\%) & F1 & Pr (\%) & Re (\%)\\

SVM & $0.52\pm 0.01$ & $\bm{68.1}\pm 0.5$ & $42.1\pm 1.0$ & $0.60\pm 0.01$ & $69.5\pm 0.66$ & $52.7\pm 0.7$ & $0.69\pm 0.01$ & $69.2\pm 1.2$ & $67.8\pm 1.2$\\
Log Reg & $0.67\pm 0.005$ & $51.9\pm 0.6$ & $\bm{93.6}\pm 2.1$ & $0.69\pm 0.01$ & $55.8\pm 1.9$ & $88.9\pm 4.0$ & $0.71\pm 0.01$ & $60.3\pm 1.4$ & $\bm{87.7}\pm 1.6$\\
LSTM & $0.67\pm 0.01$ & $54.6\pm 1.4$ & $86.9\pm 3.3$ & $0.76\pm 0.01$ & $70.6\pm 2.1$ & $82.4\pm 3.6$ & $0.79\pm 0.03$ & $74.0\pm 2.9$ & $84.6\pm 4.4$\\
HyperLSTM & $0.70\pm 0.02$ & $58.5\pm 3.2$ & $88.9\pm 3.2$ & $0.79\pm 0.04$ & $73.8\pm 5.9$ & $87.8\pm 1.6$ & $0.77\pm 0.03$ & $76.4\pm 7.9$ & $78.2\pm 2.2$\\
\hline
\mhyper{} & $\bm{0.71}\pm 0.01$ & $62.4\pm 2.9$ & $83.8\pm 5.8$ & $\bm{0.86}\pm 0.01$ & $\bm{83.9}\pm 5.1$ & $\bm{87.9}\pm 3.5$ & $\bm{0.88}\pm 0.03$ & $\bm{90.1}\pm 1.9$ & $86.3\pm 6.6$\\
\hline
\end{tabular}
\end{center}
\end{table*}
The results show that \mhyper{} achieves the highest F1-score across all window sizes. At 10s window, \mhyper{} also outperforms all baselines for precision and recall. Similar to the previous studies~\cite{liang2007,Georgiou2017,jain2016}, our results verify that that longer contextual information yield better classification accuracies across all evaluated models. The results also suggest that a trade-off between the timeliness and performance of the classification. For real-time applications, the results suggest that our method using a 10s window may offer the best trade-off.

The results suggest that LSTM-based methods are a class of flexible and expressive models capable of learning useful feature representations from gaze locations sequence and outperform the baselines that utilize handcrafted features. While it may be possible to match the performance of LSTM-based methods with more sophisticated feature engineering, the ability to automatically extract features from data is an important advantage of LSTM-based methods. The improved performance from \mhyper{} over the original HyperLSTM validates the efficacy of the proposed weight generation scheme, which uses more expressive weight generation and additional regularization.

\begin{figure}[t]
\centering
\includegraphics[width=0.38\textwidth]{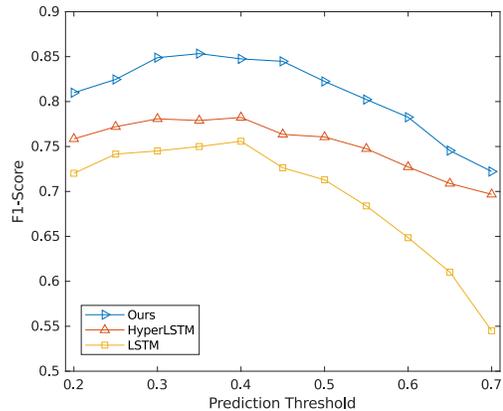}
\caption{F1-score against decision threshold for the proposed method, HyperLSTM and LSTM at $t_w=10$. The proposed method outperforms the other two and achieves a fairly consistent F1-score across increasing decision thresholds.}
\label{fig:f1}
\end{figure}

To further evaluate the performance of our model, we show in Fig. \ref{fig:f1} the impact of decision threshold on F1-score for $t_w=10 s$. Small variations in F1-score across multiple decision thresholds indicate that the model is capable of trading off between precision and recall depending on the application requirements without hurting the overall evaluation metric~\cite{jain2016}. We observe that \mhyper{} outperforms both HyperLSTM and LSTM for all the spectrum of decision thresholds, while achieving fairly consistent F1-scores across increasing decision thresholds.

\subsection{Real-Time Inference}
\mhyper{} is readily usable for real-time classification. During real-time inference, the gaze locations are aggregated into feature vectors at each second and a context of the latest $t_w$ seconds are used as input for the network to predict the current workload level. In our supplementary video, we present the real-time classification of workload for the same participants.

Real-Time classification of workload has many potential applications. As a starting point, the predicted workload could be used to manage non-critical user interaction within intelligent vehicles, such as lowering music volumes or diverting calls to voice mails to reduce workload of users~\cite{healey2005}. As model performance continue to improve, the predicted workload may be applied to more critical tasks such as deriving the control policy for human-in-loop control, as demonstrated in ~\cite{Lam2015,driggs2015}.


\section{Conclusions}
The ability to predict human cognitive states is an important problem with many applications. In this work, we addressed the problem of cognitive workload classification using a sequence of gaze locations with only consumer-grade hardware. The proposed framework is task-agnostic and generic enough for other temporal data such as EEG or ECG readings. The proposed method is able to tackle data variability commonly found in physiological signals and outperforms state-of-the-art sequence models. For future work, an interesting direction would be multi-sensory fusion, which may further improve model performance and reliability.



\section*{ACKNOWLEDGMENT}
The authors would like to thank Antoine Cully for useful discussions on this work, and all the experiment participants.
\bibliographystyle{IEEEtran}
\bibliography{ref}
\end{document}